\documentstyle[12pt]{article} 
\def\ket#1{\vert#1\rangle}
\def\bra#1{\langle#1\vert}
\def\brak#1#2{\langle#1\vert #2\rangle}

\def\bp{{\bf{p}}}
\def\bk{{\bf{k}}}
\def\bpi{{\bf{\pi}}}
\def\bxi{{\bf{\xi}}}

\def\apd{\hat a^{\dag} (\bp)}

\def\be{\begin{equation}}
\def\ee{\end{equation}}
\def\bea{\begin{eqnarray}}
\def\eea{\end{eqnarray}}
\def\dst{\displaystyle\phantom{|}}
\def\ov{\over\dst}
\def\bak{{\bf K}}
\def\dek{{\bf \Delta k}}

\normalbaselines
\begin{document}
\bibliographystyle{unsrt}

\vbox {\vspace{6mm}} 

\begin{center}
{\large \bf BROODING OVER 
PION LASERS}\\[7mm]
T. Cs\"org\H o$^{1,2}$\\
{\it $^1$ Dept. Physics, Columbia University\\
538 W 120-th Street, New York, NY 10027}\\[5mm]
J. Zim\'anyi$^2$\\
{\it $^2$ MTA KFKI RMKI \\
H - 1525 Budapest 114, POB 49, Hungary}\\[5mm]
\end{center}

\vspace{2mm}

\begin{abstract}
Brooding over pions, wave-packets and Bose-Einstein correlations,
we present a recently obtained analitical solution to a pion laser model,
which may describe the final state of pions in high energy heavy ion
collisions.
\end{abstract}

\underline{\it Introduction.}
 At the planned RHIC accelerator pions, the lightest observable strongly
interacting particles, will be produced in thousands in a unit
interval of the longitudinal phase-space. At such high multiplicities,
the bosonic nature of the pions and the resulting Bose-Einstein
statistics  may dominate the behaviour of the observable physical
quantities, like the multiplicity distribution, the single-particle
momentum spectra and the second order intensity correlation function
(frequently referred to as the Bose-Einstein correlation function or
BECF). Present experiments at CERN SPS  involving central collisions of
two almost completely stripped nuclei (heavy ions) already
produce cca 600 similar pions in a unit value of rapidity,
when $Pb + Pb$ reactions are measured at 160 AGeV laboratory
bombarding energy. At the RHIC collider, to be completed
by 1999, thousands of pions
could be produced in a unit rapidity interval in a 40 TeV
collision of $Au$ nuclei~\cite{qm96,s96}.
This multiplicity is large enough for the higher order Bose-Einstein
symmetrizations to reveal their effects on the observables.
We report here about a simple model~\cite{pwlaser,prclaser}
that is analytically solvable
both in the dense and the rare Bose-gas limit. The model is based on
stimulated emission of pion wave-packets that is induced by the
overlap of the multi-pion wave-packet states.
The presentation closely follows the lines of ref.~\cite{pwlaser}.

\underline{\it Specification of the Pion-Laser Model.}
If the number of pions
in a unit value of phase-space is large enough these bosons may condense
into the same quantum state and a pion laser could be created~\cite{plaser}.
Similarly to this process, when a large number of bosonic atoms
are collected in a magnetic trap and cooled down to increase their
density in phase-space, the bosonic nature of the atoms reveals itself
in the formation of a Bose-Einstein condensate~\cite{atom},
a macroscopic quantum state. Such a condensation mechanism
may provide the key to the formation of atomic lasers in condensed matter
physics and to the formation of pion lasers in high energy particle and
heavy ion physics, reviewed recently in refs~\cite{bengt,bill}.

The density matrix of a generic quantum mechanical
system is
\be
\hat \rho = \sum_{n=0}^{\infty} \, {p}_n \, \hat \rho_n,
\ee
Here the index $n$ characterizes sub-systems with
particle number fixed to $n$,
the multiplicity distribution is prescribed by
the set of $\left\{ {p}_n\right\}_{n=0}^{\infty}$, normalized as
$\sum_{n= 0}^{\infty} p_n = 1$.
The density matrixes are normalized as
$\mbox{\bf Tr} \, \hat \rho = 1$ and $\mbox{\bf Tr} \, \hat \rho_n = 1$,
where
\be
\hat \rho_n \!\! = \!\! \int \!\! d\alpha_1 ... d\alpha_n
\,\,\rho_n(\alpha_1,...,\alpha_n)
\,\ket{\alpha_1,...,\alpha_n} \bra{\alpha_1,...,\alpha_n}
\ee
and the states $\ket{\alpha_1,...,\alpha_n}$ denote properly normalized
$n$-particle wave-packet boson states.

A wave packet creation operator is
\bea
\alpha_i^{\dag} \! & = &\int\!\! {d^3\bp \over
(\pi \sigma^2 )^{3\over 4} } \!
	\mbox{\rm e}^{\! -{( \bp- \bpi_i)^{2}\over 2 \sigma_i^{2}}
	-  i \bxi_i (\bp - \bpi_i) +
 	 i \omega(\bp) ( t-t_i) } \ \apd ,
\label{e:4}
\eea
where $\alpha_i = (\bxi_i, \bpi_i, \sigma_i,t_i)$ refers
to the parameters of the
wave packet $i$: the center in space, in momentum space, the width
in momentum space and the production time, respectively.
For simplicity we assume that all the wave packets are
emitted at the same instant and with the same width,
$\alpha_i = (\bpi_i, \bxi_i, \sigma, t_0)$.
The $n$ boson states, normalized to unity, are given as
\be
\ket{\ \alpha_1, \ ...\ , \ \alpha_n} =
 {\left({ \displaystyle{\strut \sum_{\sigma^{(n)} }
	\prod_{i=1}^n \brak{\alpha_i}{\alpha_{\sigma_i}} } }
	\right)^{- {1\over 2}} } \!\!
\  \alpha^{\dag}_n  \ ... \
\alpha_1^{\dag} \ket{0}.
\label{e:expec2}
\ee
Here $\sigma^{(n)}$ denotes the set of all the permutations of
the indexes $\left\{1, 2, ..., n\right\}$
and the subscript sized $_{\sigma_i}$ denotes the index that
replaces the index $_i$ in a given permutation from $\sigma^{(n)}$.

\underline{\it Solution for a New Type of Density Matrix \label{s:3}}
There is one special density matrix, for which one can
overcome the difficulty, related to the
non-vanishing overlap of many hundreds of wave-packets,
even in an explicit analytical manner.
Namely, if one assumes, that we have
a system, in which the emission probability  of a boson is increased if
there is an other emission in the vicinity:
\be
\rho_n(\alpha_1,...,\alpha_n) \! = \!
	{\dst 1 \ov {\cal N}{(n)}} \!
 \prod_{i=1}^n \rho_1(\alpha_i)  \!
\left(\sum_{\sigma^{(n)}} \prod_{k=1}^n \,
\brak{\alpha_k}{\alpha_{\sigma_k}}
\right)
\label{e:dtrick}
\ee
The coefficient of proportionality, ${\cal N}{(n)}$,
 can be determined from the normalization condition.
 The density matrix of eq.~(\ref{e:dtrick}) describes a
 quantum-mechanical wave-packet system with induced emission, and the
 amount of the induced emission is controlled by the overlap of the $n$
 wave-packets~\cite{prclaser}, yielding a weight in the range of $[1,n!]$.
 Although it is very difficult numerically to operate with such a
 wildly fluctuating weight, we were able to reduce the problem~\cite{prclaser}
 to an already discovered ``ring" - algebra of permanents for plane-wave
 outgoing states~\cite{plaser}.

For the sake of simplicity we assume
a non-relativistic, non-expanding
static source at rest in the frame where the calculations are performed.
This may correspond to the central, mid-rapidity piece of the
highly relativistic fireball created at RHIC.
\bea
\rho_1(\alpha)& = &\rho_x(\bxi)\, \rho_p (\bpi)\, \delta(t-t_0), \cr
\rho_x(\bxi) & = &{1 \over (2 \pi R^2)^{3\over 2} }\, \exp(-\bxi^2/(2
R^2) ), \cr
\rho_p(\bpi) & = &{1 \over (2 \pi m T)^{3\over 2} }\, \exp(-\bpi^2/(2 m
T) ),
\eea
and a Poisson multiplicity distribution $p_n^{(0)}$ for the case when
the Bose-Einstein effects are negligible:
\be
        {p}^{(0)}_n = {n_0 ^n \over n!} \exp(-n_0),
\ee
 This prescribes the multiplicity distribution
of pion lasers in a very rare gas limit~\cite{plaser},
and completes the specification of the model.
The plane-wave model, to which the multi-particle
wave-packet model was reduced in ref.~\cite{prclaser},
 can be further simplified~\cite{prclaser} to a set of
recurrence relations with the help
of the so-called ``ring-algebra" discovered first
by S. Pratt in ref.~\cite{plaser}.
The probability of finding events with multiplicity $n$,
as well as the single-particle and the two-particle momentum distribution in
such  events is given as
\bea
	{p}_n  & = &
		\omega_{n} \left( \sum_{k=0}^{\infty} \omega_{k} \right)^{-1},
			\label{e:d.1}\\
	N^{(n)}_1(\bk_1)  &  = &
                \sum_{i=1}^n  {\dst  \omega_{n-i} \ov \omega_{n}}
		 G_i (1,1) , \label{e:d.2} \\
	N^{(n)}_2(\bk_1,\bk_2)   &  =  &
		\sum_{l=2}^n
		\sum_{m=1}^{l-1}
		{\dst \omega_{n-l}  \ov \omega_n}
		\left[ G_m(1,1) G_{l-m} (2,2) +
		 G_m(1,2) G_{l-m}(2,1)\right]
		, \label{e:d.3}
\eea
where $\omega_n = p_n / p_0$ and
\be
        G_n(i,j)  =
		n_0^n \, h_n \exp(-a_n (\bk^2_i + \bk^2_j) + g_n \bk_i \bk_j).
                        \label{e:3.7}
			\label{e:a.1}
\ee
Averaging over the multiplicity distribution $p_n$ yields the inclusive
spectra as
\bea
	G(1,2) & = & \sum_{n=1}^{\infty} G_n(1,2), \\
	N_1(\bk_1) & = & \sum_{n=1}^{\infty} p_n N^{(n)}_1(\bk_1)
		\, = \,  G(1,1), \\
	N_2(\bk_1,\bk_2) & = & G(1,1) G(2,2) + G(1,2) G(2,1).
\eea
With the help of the notation
\bea
        \sigma_T^2  =  \sigma^2 + 2 m T,
	\label{e:sigt}
	& \qquad &
        R_{e}^2  =  R^2 + {\dst m T \ov \sigma^2 \sigma_T^2},  \label{e:reff}
\eea
the recurrence relations that
correspond to the solution
of the ring-algebra~\cite{chao,plaser}
are obtained~\cite{prclaser}
for the case of the multi-particle wave-packet model. These
correspond to the pion laser model of
S. Pratt when a replacement $ R \rightarrow R_{e}$
and $T \rightarrow T_{e} = \sigma_T^2/(2 m)$ is performed.

Let us introduce the following auxiliary quantities:
 \bea
	\gamma_{\pm}
		 =  {\dst 1\ov 2} \left( 1 + x \pm \sqrt{1 + 2 x} \right)
			\label{e:gam.s} & \qquad  &
	x  =  R_{e}^2 \sigma_T^2
			\label{e:di.x}
 \eea
	The {\it general analytical solution} for the multiplicity
	distribution of the model is given through
	the generating function of the multiplicity distribution $p_n$
 \bea
	G(z) & = & \sum_{n = 0}^{\infty} p_n z^n  \label{e:d.g}
	\, = \, \exp\left( \sum_{n=1}^{\infty} C_n (z^n - 1) \right),
	\label{e:gsolu}
 \eea
	where $C_n$ is
 \be
	 C_n  =  {\dst n_0^n \ov n}
			\left[ \gamma_+^{n\over 2} -
				\gamma_-^{n\over 2} \right]^{-3 },
		\label{e:c.s}
 \ee
	together with the {\it general analytic solution}
	for the functions $G_n(1,2)$:
 \bea
	G_n(1,2)\! & = & \! j_n
	\mbox{\rm e}^{
	- {b_n \over 2} \left[
		\left(\gamma_+^{n\over 2} \bk_1
		- \gamma_-^{n \over 2} \bk_2\right)^2
		+ \left(\gamma_+^{n\over 2} \bk_2 -
		  \gamma_-^{n \over 2} \bk_1\right)^2 \right]
	},\!\! \\
	j_n \! & = & \! n_0^n\! \left[{ b_n \over \pi}\right]^{3 \over 2}
	\qquad
	b_n \,  = \, {\dst 1 \over  \sigma_T^2}
			{\dst \gamma_+ - \gamma_- \ov \gamma_+^n - \gamma_-^n}
 \eea
The detailed proof that
 the analytic solution to the  multi-particle wave-packet  model is indeed
given by the above equations is described in ref.~\cite{prclaser}.
	The representation of eq.~(\ref{e:gsolu}) indicates that
	the quantities $C_n$-s are the so called combinants
	~\cite{gyul-comb,hegyi-c1,hegyi-c2} of the
	probability distribution of $p_n$ and in our case their
	explicit form is known for any set of model parameters, as
	given by eqs.
	~(\ref{e:c.s},\ref{e:di.x},\ref{e:reff}).
	One can prove~\cite{prclaser}, that the mean multiplicity
	 $ \langle n \rangle =
	\sum_{n=1}^{\infty} n p_n =  \sum_{i = 1}^{\infty} i C_i$.
	The large $n$ behavior of $n C_n $ depends on the ratio of
	$ n_0 / \gamma_+^{3\over 2}$, since for large values of $n$,
	we always have $ (\gamma_- / \gamma_+)^{n\over 2} << 1$.
	The critical value of $n_0$ is
 \bea
	n_c & = & \gamma_+^{3\over 2} = \left[ {\dst 1 + x + \sqrt{1 + 2 x}
			\ov 2 } \right]^{3\over 2}.
 \eea
	If $n_0 < n_c$,
	one finds $lim_{n \rightarrow \infty} n C_n = 0$
	and $\langle n \rangle < \infty$,
	if  $n_0 >  n_c$
	one obtains $lim_{n \rightarrow \infty} n C_n = \infty $
	and $\langle n \rangle = \infty$,
	finally,
	if $n_0 = n_c$ one finds
	$lim_{n \rightarrow \infty} n C_n = 1$
	and $\langle n \rangle = \infty$.
	The divergence of the mean multiplicity
	$ \langle n \rangle $ is related to condensation of the
	wave-packets to the wave-packet state with $\bpi = 0$,
	i.e. to the wave-packet state with zero mean momentum~\cite{prclaser},
	if $n_0 \ge n_c$.

	The multiplicity distribution of eq.~(\ref{e:gsolu}) is
	studied at greater length in ref.~\cite{prclaser}.

\underline{\it Dense Bose-gas Limit:}
	This wave-packet model exhibits a lasing behavior in the very dense
	Bose-gas limit, which corresponds to an optically coherent
	behavior, characterized by a vanishing enhancement of the
	two-particle intensity correlations at low momentum,
	$C(\bk_1,\bk_2) = 1$, a case which
	is described in greater details in ref.~\cite{prclaser}.

 \underline{\it Rare gas limiting case}.
 Large source sizes or
 large effective temperatures correspond to
 the $ x >> 1 $ limiting case,
 where the general analytical solution of the model, presented above,
 becomes particularly simple and the exclusive and inclusive spectra
 and correlation functions can be obtained analytically
	to leading order in $ 1/x << 1$.
 From eq.~(\ref{e:a.1}) one obtains that
 \bea
	G_n(1,2) & = &  j_n
		\exp\!\!\left[
		 - {\dst n \ov 2 \sigma_T^2 }( \bk_1^2 + \bk_2^2)
		- \! {R_{e}^2 \ov 2 n}
			{\bf \Delta}\bk^2 \right]\!, \\
	\label{e:g.i.rare}
	j_n & = &{\dst n^{5/2} C_n \ov (\pi \sigma_T^2)^{3\over 2} },
	\qquad
	C_n \,\, = \,\, { \dst n_0^n  \ov n^4 }
		\left({\dst 2 \ov x} \right)^{{3\over 2}(n-1)}\!\!,
	\label{e:c.rare}
 \eea
	where 	${\bf \Delta}\bk = \bk_1 - \bk_2$.
	We can see from eq.~(\ref{e:g.i.rare}), that
	the higher order corrections will contribute to the observables
	with reduced effective temperatures and reduced effective radii.

	The {\it very} rare gas limiting case  corresponds to keeping only the
	leading $n=1$ order terms in the above equations.
	The multiplicity distribution
	is a Poisson distribution with
	$\langle n \rangle = n_0$ and no influence from stimulated emission.
	The momentum distribution is a Boltzmann
	distribution, and the exclusive and inclusive momentum distributions
	coincide. The leading order two-particle Bose-Einstein
	correlation function is a Gaussian correlation function
	with a constant intercept parameter of $\lambda = 1$ and
	with a momentum - independent radius parameter of $R_* = R_{e}$
	~\cite{prclaser}.

	The two-particle exclusive
correlation functions can also be evaluated by
applying a Gaussian approximation to the
leading order corrections in the $x >> 1$ limiting case:
\bea
C^{(n)}_2(\bk_1,\bk_2) \!\! & = &   {\dst n^2 \ov n(n-1) }
		{\dst N_2^{(n)}(\bk_1,\bk_2) \ov
		N_1^{(n)}(\bk_1) \, N_1^{(n)}(\bk_2) }
	\,	= \,
		1 + \lambda_{\bak} \!
		\exp\!\left( - R_{\bak,s}^2  \dek_{s}^2
		- R_{\bak,o}^2 \dek_{o}^2 \right), \!
\eea
where $\bak = 0.5 (\bk_1 + \bk_2)$, the side and outwards directions
are introduced utilizing the spherical symmetry of the source as
$\dek_{s} = \dek - \bak{ (\dek \cdot \bak) / ( \bak \cdot \bak)} $ and
$\dek_{o} = \bak{ (\dek \cdot \bak) / ( \bak \cdot \bak)} $,
 similarly to refs.~\cite{nr}. The mean-momentum dependent intercept
and radius parameters are
\bea
\lambda_{\bf{K}} \!\! & = & \! \!  1 +
		{\dst 2 \ov ( 2 x)^{3\over 2} }
		\left[ 1 -
		2^{(5/2)} \exp\left( - {\dst {\bf K}^2 \ov \sigma_T^2 } \right)
		\right] \label{e:lambda.corr} \\
R_{{\bf{K}},s}^2 \!\! & =   &\!\!  R_{e}^2  +
		{\dst 1 \ov ( 2 x)^{3\over 2} }
	 	\left[  R_{e}^2 - \sqrt{2}
		\exp\left( - {{\bf K}^2 \ov \sigma_T^2} \right)
		\left(
		 ( n + 2) R_{e}^2 + {\dst 2 \ov \sigma_T^2 }
		\right)
		\right], \label{e:r.corr}\\
R_{{\bf K},o}^2 \!\! & = & \!\! R_{{\bf{K}},s}^2 +
		{\dst n \ov    x^{3\over 2} } {\dst {\bf K}^2  \ov \sigma_T^4 }
		\exp\left( - {\dst {\bf K}^2 \ov \sigma_T^2} \right)
		\label{e:c.dir}
\eea
Thus the symmetrization results in a momentum - dependent
intercept parameter $\lambda_{\bak}$ that starts from a
$\lambda_{\bak = 0} < 1 $ value at low momentum  and {\it increases}
with increasing momentum. Already in the first paper about the pion laser
model, ref.~\cite{plaser}, a reduction of the exact intercept parameter
was observed and interpreted as the onset of a coherent behavior in the
low momentum modes. First, a partially coherent system is created,
characterized by $\lambda_{\bak} < 1$, and if the density of pions
is further increased, one finds  a fully developed pion-laser
with $\lambda_{\bak} = 0$,
see ref.~\cite{prclaser} for analytic considerations.

Observe that the radius parameter at low mean momentum
decreases while the radius parameter at high mean momentum
increases, as compared to $R_{e}$.
{\it
The radius parameter of the exclusive correlation function thus becomes
momentum-dependent even for static sources!}
This  effect is more pronounced for higher values of the
fixed multiplicity $n$, in contrast to the momentum dependence of
$\lambda_{\bak}$ that is independent of $n$.

Last, but not least, a specific term appears in the two-particle
exclusive correlation function, that contributes only to the {\it
out} direction, which, in case of spherically symmetric sources,
may be identified with the direction of the mean momentum~\cite{nr}.
This directional dependence is related only to the direction of the relative
momentum as compared to the direction of the mean momentum,
and does not violate the assumed spherical symmetry of the boson source.
The effect vanishes both at very low or at very high values of the mean
momentum
${\bf K}$, according to eq.~(\ref{e:c.dir}).

\underline{\it Highlights :}
Our new analytic result is
that multi-boson correlations
generate  {\it momentum-dependent}
radius and intercept parameters even for {\it static} sources, as well
as induce a special {\it directional dependence} of the correlation
function. The effective {\it radius parameter}  of the two-particle
correlation function {\it is reduced  for low values }
and {\it enlargened for large values of the mean momentum}
in the rare gas limiting case, as compared to the case
when multi-particle symmetrization effects
are neglected.
 From the numerical studies,
described in ref.~\cite{pwlaser},
we find that for an extended, hot and rare gas of a few hundred pions, the
reduction of the radius parameters at low momentum  is found to
be the most apparent effect. The directional dependence of the
radius parameters and the enhancement of the radii at high momentum
is characteristic for a small, cold pion gas with only a handful of
particles in it.
These results can be understood qualitatively by an enhancement
of the wave-packets in the low momentum modes,
due to multi-particle Bose-Einstein symmetrization effects,
as the system starts to approach the formation of a laser,
characterized by the appearance of
partial optical coherence in the low momentum modes.

\section*{Acknowledgments:}
Cs. T. would like to thank M. Gyu\-lassy, S. Hegyi, G. Vahtang  and X. N. Wang
for stimulating discussions, to J. Janszky and T. Kiss for a very
stimulating conference athmosphere at Balatonf\"ured and for support.
This work was supported by the grants OTKA
 W01015107, T016206 and T024094, MAKA 378/93 and
by an Advanced Research Award from the Fulbright Foundation.

\setlength{\itemsep}{0pt}

\end{document}